\documentclass[printer]{aa}  

\usepackage{graphicx}
\usepackage{txfonts}
\usepackage{hyperref}
\usepackage{verbatim}
 \usepackage{longtable,lscape}
 \usepackage{epsfig}
\usepackage{natbib}
\usepackage{setspace}\usepackage{threeparttable}
\newcommand{\RM}[1]{\MakeUppercase{\romannumeral #1}}
\bibpunct{(}{)}{;}{a}{}{,} 
\def\fss{\hbox{$.\!\!^{\rm s}$}}        

\begin{document}
   \title{\textit{Fermi}/GBM observations of the ultra-long GRB~091024:}

   \subtitle{A burst with an optical flash.}

   \author{D.Gruber\inst{1},
		T. Kr\"uhler\inst{1,2},
		S. Foley\inst{1},
		M. Nardini\inst{1},
		D. Burlon\inst{1},
		A. Rau\inst{1},
		E. Bissaldi\inst{1},
		A. von Kienlin\inst{1},
		S. McBreen\inst{1,3},
		J. Greiner\inst{1},
		P. N. Bhat\inst{4},
		M. S. Briggs\inst{4},
		J. M. Burgess\inst{4},
		V. L. Chaplin\inst{4},
		V. Connaughton\inst{4},
		R. Diehl\inst{1},
		G. J. Fishman\inst{5},
		M. H. Gibby\inst{6},
		M. M. Giles\inst{6},
		A. Goldstein\inst{4},
		S. Guiriec\inst{4},
		A. J. van der Horst\inst{4},
		R. M. Kippen\inst{7},
		C. Kouveliotou\inst{5},
		L. Lin\inst{4},
		C. A. Meegan\inst{8},
		W. S. Paciesas\inst{4},
		R. D. Preece\inst{4},
		D. Tierney\inst{3},
		C. Wilson-Hodge\inst{5}.
          }

   \institute{Max Planck Institute for extraterrestrial Physics, 
   		Giessenbachstrasse, Postfach 1312, D-85748, Garching, Germany\\
		\email{dgruber@mpe.mpg.de}
		\and
		 Universe Cluster, Technische Universit\"at M\"unchen, 
		 Boltzmannstra\ss e 2, D-85748, Garching, Germany
		\and
		University College, Dublin, Belfield, Stillorgan Road, Dublin 4, Ireland
		\and
		University of Alabama in Huntsville, NSSTC, 
		320 Sparkman Drive, Huntsville, AL 35805, USA 
		\and
		Space Science Office, VP62, NASA/Marshall Space Flight Center
		Huntsville, AL 35812, USA
		\and
		Jacobs Technology, Inc., Huntsville, Alabama
		\and
		 Los Alamos National Laboratory, 
		 P.O. Box 1663, Los Alamos, NM 87545, USA
		 \and
		 Universities Space Research Association, 
		 NSSTC, 320 Sparkman Drive, Huntsville, AL 35805, USA
		  }
   \date{}

 
  \abstract
   {}
   {In this paper we examine gamma-ray and optical data of GRB~091024, a gamma-ray burst (GRB) with an extremely long duration of $T_{\rm{90}}\approx1020$~s, as observed with the \textit{Fermi} Gamma-Ray Burst Monitor (GBM). }
   {We present spectral analysis of all three distinct emission episodes using data from \textit{Fermi}/GBM. Because of the long nature of this event, many ground-based optical telescopes slewed to its location within a few minutes and thus were able to observe the GRB during its active period. We compare the optical and gamma-ray light curves. Furthermore, we estimate a lower limit on the bulk Lorentz factor from the variability and spectrum of the GBM light curve and compare it with that obtained from the peak time of the forward shock of the optical afterglow.
   } 
   {From the spectral analysis we note that, despite its unusually long duration, this burst is similar to other long GRBs, i.e. there is spectral evolution (both the peak energy and the spectral index vary with time) and spectral lags are measured.
   We find that the optical light curve is highly anti-correlated to the prompt gamma-ray emission, with the optical emission reaching the maximum during an epoch of quiescence in the prompt emission. We interpret this behavior as the reverse shock (optical flash), expected in the internal-external shock model of GRB emission but observed only in a handful of GRBs so far. 
   The lower limit on the initial Lorentz factor deduced from the variability time scale ($\Gamma_{min}=195_{-110}^+{90}$)is consistent within the error to the one obtained using the peak time of the forward shock ($\Gamma_0=120$) and is also consistent with Lorentz factors of other long GRBs.}
   {}


  \authorrunning{D. Gruber et al.}
   \maketitle
%

\section{Introduction}
The \textit{Fermi} spacecraft was successfully launched on June 11, 2008. Its payload includes the Large Area Telescope \citep[LAT; ][]{atwood09} and the Gamma -Ray Burst Monitor \citep[GBM; ][]{meegan09}. Specifically designed for Gamma-Ray Burst (GRB) studies, GBM observes the whole unocculted sky with a total of 12 sodium iodide scintillation detectors (NaI), sensitive between 8 keV to 1 MeV,  and two bismuth germanate (BGO) detectors, covering an energy range from 200 keV to 40 MeV. In this way, GBM offers a novel view of GRBs, the most energetic events in the Universe after the Big Bang.
Since their detection in the late 60s \citep{kleb73}, major efforts have been undertaken to solve this enigmatic puzzle and it is now established that they originate from highly relativistic collimated outflows from a compact source with Lorentz factors $\Gamma > 100$. However, to date, major aspects of GRBs (the mechanism creating gamma-rays, the content of the jet, the magnetic field, etc.) are not well understood.

In this paper we present the gamma-ray observations together with already published optical observations of GRB~091024, a very long burst with a duration of approximately 1020~s.
GBM triggered and located GRB~091024 at  08:55:58.47 ($t_0$) and triggered a second time at 09:06:29.36 UT \citep{bis09}. GRB~091024 was also seen by Konus-Wind \citep{gol09}, SPI-ACS (Rau, priv. comm.) and \textit{Swift} \citep{mar09}. However the burst was outside the \textit{Swift}-BAT Field of view (FOV) after $t_0+460$~s due to an Earth-limb constraint. \textit{Swift}-XRT determined the position to be $\alpha_{J2000}=22^h 37^m 00\fss4$ and $\delta_{J2000}= 56^\circ 53' 21''$ with an uncertainty of 6 arcsec \citep{page09}. Unfortunately, XRT started observing the field of GRB~091024 about 53 minutes after the BAT trigger. 
\textit{Fermi} entered the South Atlantic Anomaly (SAA) 2830~s after  $t_0$, by which time GRB emission cannot be distinguished from the background.
An autonomous repoint request (ARR) was issued by \textit{Fermi} in order to align the burst with the FoV of the LAT at 09:12:14.28 UT, i.e. $976$~s after the first trigger. However, no significant emission was detected in the LAT energy range during any of the time intervals in which the burst was in the LAT field of view \citep{bouvier09}.

Not many GRBs have been observed in the optical band while the prompt gamma-ray emission was still active \citep[the best example being GRB~990123 and GRB~080319B, see e.g.][]{akerlof99,  racusin08}. For GRB~091024 optical data were acquired soon after the first trigger, and throughout its active phase.  \citet{henden09} obtained photometry using the Sonoita Research Observatory (SRO) starting 540~s after the trigger. Ten $R_c$-band, nine $V$-band, and one $I_c$-band exposures were acquired.
The 2m - Faulkes Telescope North started observing the field of GRB~091024 207~s after the trigger \citep{cano09}. The 0.6m Super LOTIS telescope started observing 58~s after the BAT trigger \citep{updike09}.

Optical spectra of the afterglow were obtained with the Low Resolution Imaging Spectrometer mounted on the 10-m Keck I telescope and the GMOS-N spectrograph at Gemini North, 
revealing a redshift of $z=1.09$ \citep{cenko09, cucc09}

This paper is organized as follows. In Sect. \ref{sec:datana} we describe the GBM observation and data reduction together with the spectral and spectral lag analysis of the three well-defined emission episodes. In Sect. \ref{sec:afterglow} we describe the behavior of the optical afterglow data compared to the prompt gamma-ray emission. In Sect. \ref{sec:lorentz} we estimate a lower limit on the initial Lorentz factor using the variability time scale of the prompt emission. We discuss the position of GRB~091024 in the Yonetoku- and Amati relations in Sect. \ref{sec:relations}. Finally, in Sect. \ref{sec:sum} we summarize our findings and conclusions.

Throughout this paper we use a flat cosmology with $\Omega_m=0.32$, $\Omega_{\Lambda}=0.68$ and $H_0=72 \rm{km} \rm{s}^{-1} \rm{Mpc}^{-1}$ \citep{bennet03, spergel03}.

\section{GBM data analysis}
\label{sec:datana}
\begin{figure}
\centering
\includegraphics[width=0.5\textwidth]{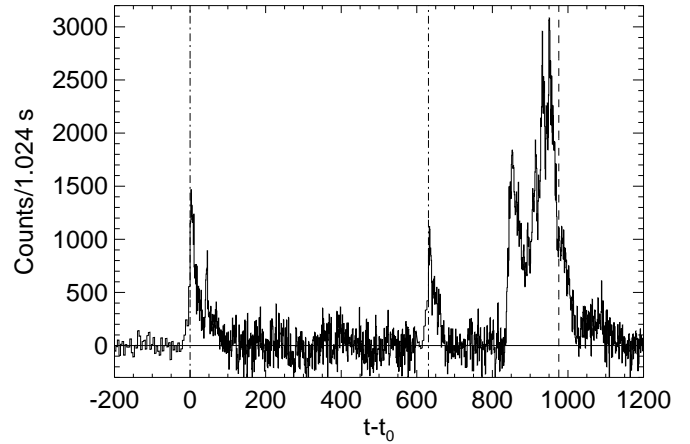}%
\caption{Background corrected light curve of GRB091024 in the energy range 8 - 1000 keV. The vertical dash-dotted lines show the times of the two triggers and the dashed line the beginning of the ARR.}
\label{fig:lc}
\end{figure}

Using all 12 NaI detectors, we created the background subtracted light curve shown in Fig.~\ref{fig:lc}. It shows three distinct emission periods, separated by two periods of quiescence. The first emission episode consists of at least two FRED (Fast Rise, Exponential Decay) like pulses (hereafter episode \RM{1}). The time in which 90\% of the fluence is observed is $\rm{T}_{90,\rm{I}} = 72.6 \pm 1.8\,\rm{s}$. Another emission episode starts 630~s after $t_0$. This emission period, which actually triggered GBM a second time, consists of an initial pulse ($\mathrm{T}_{90,\rm{II}}=44.5 \pm 5.4\,\rm{s}$, hereafter episode \RM{2}). A multi-peaked episode starts 200~s later (corresponding to 840~s after $t_0$) and continues for 350~s with $T_{\rm{90,III}}=150 \pm 10\,\rm{s}$ (hereafter episode \RM{3}). 

Due to the highly variable background caused by \textit{Fermi}'s ``rocking angle'' observing mode, it is impossible to detect an excess in count rate during the two epochs between episode \RM{1} and \RM{2} and between \RM{2} and \RM{3} in the GBM data. We conclude that the GRB signal during these intervals, if any, is below background level. Thus, we define these two epochs as phases of quiescence.

With its very long duration of $T_{\rm{90}}\approx1020$~s, GRB~091024 is the longest burst detected by \textit{Fermi}/GBM and also one of the longest bursts ever seen (see Table \ref{tab:longbursts}). Among the longest events, only GRB~020410 and GRB~970315D show a multi-peaked behavior whereas the others have a long lasting FRED-like pulse.

\begin{table*}[ht]
\renewcommand{\arraystretch}{1.3}
\caption[ ]{Longest bursts known to date. $T_{\rm{90}}$ of GRB~970315D was taken from the current BATSE 5B catalogue \url{http://gammaray.nsstc.nasa.gov}.}
\label{tab:longbursts}
\begin{center}
\begin{tabular}{l r c c c }
\hline
	 & $T_{90} [s]$& z & Observatories & Refs. \\
\hline
\hline											
GRB~060814B  & $\approx 2700$& 0.84 & Konus-Wind & \citet{palshin08} \\ 
GRB~971208    & $\approx 2500$& - & \textit{BATSE}, Konus-Wind &  \citet{giblin02, palshin08} \\ 
GRB~060218 &  $2100$& 0.033 & \textit{Swift}-BAT &  \citet{campana06} \\ 
GRB~020410 & $\approx 1550$& - & $Beppo$SAX, Konus-Wind &  \citet{nicastro04} \\ 
GRB~970315D & $1307$& & \textit{BATSE} & unpublished (see caption) \\ 
GRB~091024 & 1020 & 1.09 & \textit{Fermi}/GBM, Konus-Wind & this paper \\
\hline
\end{tabular}
\end{center}
\smallskip 

\end{table*}

\subsection{Spectral analysis}
Photons are detected up to $\sim 500$~keV during all three emission episodes. 
For the spectral analysis we determined which of the GBM detectors had source angles $\epsilon \le 60^\circ$ for the whole duration of the three distinct emission periods and, at the same time, were not occulted either by the spacecraft or by the solar panels. Only data from detectors fulfilling these criteria (see Table \ref{tab:fitvalues}) were used for the analyses. Even though both BGO detectors show little signal throughout the duration of the burst, they were included nonetheless for the spectral analysis to get an upper limit at high energies.

For the purposes of our spectral analysis we used CSPEC data \citep{meegan09}, from 8~keV to 40~MeV, with a temporal resolution of 1.024~s. For each emission episode we fitted low-order polynomials to a user defined background interval before and after the prompt emission for every energy channel and interpolated this fit across the source interval. The spectral analysis was performed with the software package RMFIT (version 3.3rc8) and the GBM Response Matrices v1.8. 

Three model fits were applied, a single power-law (PL), a power-law function with an exponential high-energy cutoff (COMP) and the Band function \citep{band93}. The best model fit is the function which provides the lowest Castor C-stat\footnote{http://heasarc.nasa.gov/lheasoft/xanadu/xspec/manual/\\XSappendixCash.html} value (Cash 1979). The profile of the Cash statistics was used to estimate the $1\sigma$ asymmetric error.

\subsubsection*{Emission episodes \RM{1} and \RM{2}.}

Detectors NaI~7 ($53^\circ$), NaI~8 ($8^\circ$) and NaI~11 ($55^\circ$) had an unobstructed view of episode \RM{1}. Although the GRB illuminated the spacecraft from the side, thus illuminating the BGOs through the photomultipliers, we included BGO~1 for the spectral analysis since the detector response matrix (DRM) accounts for this effect. The spectral fit was performed over the $\rm{T}_{90,I}$ interval, i.e. from -3.8~s to 67.8~s. The COMP model, with $E_{peak}=412^{+69}_{-53}$~keV and energy index $-0.92 \pm 0.07$  provides the best fit to the data. 

Episode \RM{2} shows a single emission period. Different detectors, i.e. NaI~6 ($27^\circ$), NaI~7 ($50^\circ$) and NaI~9 ($32^\circ$) and BGO~1, fulfilled the selection criteria and were used for the spectral analysis. A COMP model with $E_{peak}=371^{+111}_{-71}$~keV and an index of $-1.17 \pm 0.07$ fits episode \RM{2} best. 

The spectral parameters of both precursors are listed in Table~\ref{tab:fitvalues}.

\subsubsection*{Episode III}
\begin{figure}
\centering
\includegraphics[width=0.35\textwidth, angle=90]{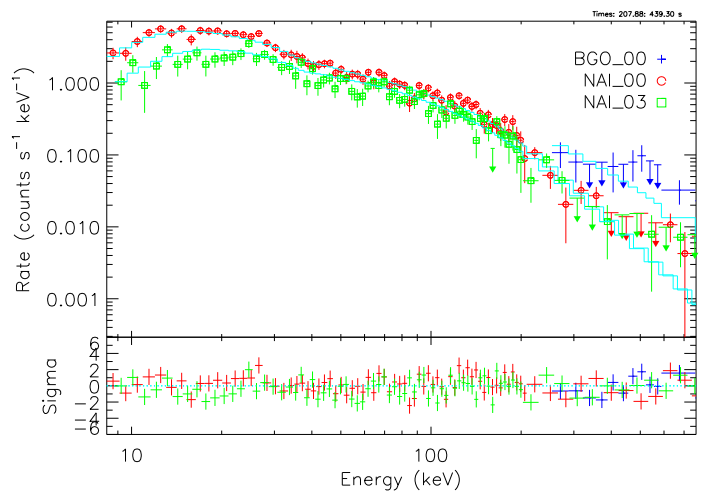}
\caption{Count spectrum of episode \RM{3}, best fit by a COMP model with index $\alpha=-1.38 \pm 0.02$ and a high-exponential cutoff located at $E_{peak}=278^{+22}_{-18}$~keV.}
\label{fig:specmainem}
\end{figure}

For episode \RM{3}, starting around $830$~s after trigger time, NaI~0 ($13^\circ$), NaI~1 ($36^\circ$), NaI~3 ($53^\circ$) and BGO~0 are not occulted by the spacecraft. Due to the location of the NaI detectors on the spacecraft, detector NaI~0 blocks detector NaI~1, resulting in a significant reduction of effective area. Therefore we excluded NaI~1 from the spectral analysis.
Episode \RM{3} is best fit by a COMP model with index $\alpha=-1.38 \pm 0.02$ and an exponential high energy cutoff located at $E_{peak}=278^{+22}_{-18}$~keV (see Fig.~\ref{fig:specmainem}).

There is no indication of the first emission episodes to be systematically softer than episode \RM{3}. If one adopts the definition of precursors as in \citet{burlon08}, we conclude that episodes \RM{1} and \RM{2} are likely to be produced by the same engine that produced the main GRB emission \citep[for details see][]{burlon08}. 

In all cases, $E_{peak}$ is at about the upper end of the detected photon energies. This explains why the high energy index $\beta$ of the Band function is only poorly constrained.

Fig.~\ref{fig:HR} shows the evolution of the Hardness Ratio (HR), defined as the ratio of the energy flux in the 100 keV - 300 keV and 10 keV - 100 keV range along the duration of the burst. There is a hard-to-soft evolution which is represented also in the evolution from higher to lower $E_{peak}$ of the three emission episodes. A similar behavior is seen in other long GRBs \citep{kocevski:2003,hafizi:2007}. It is interesting to note that the overall HR-evolution over the burst duration is much larger than the canonical HR-intensity correlation between the 3 emission episodes - see the particularly soft HR of episode \RM{3} despite its high intensity.

\begin{table*}[ht]
\renewcommand{\arraystretch}{1.3}
\caption[ ]{Best fitting spectral parameters for the three emission episodes in GRB~091024.}
\label{tab:fitvalues}
\begin{center}
\begin{tabular}{cccccccc}
\hline
	T-T$_0$  & model & $E_p$ & $\alpha$ / index & C-Stat/DOF & Fluence\\
	 $[\mathrm{s}]$ &  & $ [\mathrm{keV}]$ &     &   &   [$10^{-5}\,\mathrm{erg}/\mathrm{cm}^2$]\\
\hline
\hline											
	-3.8 : 67.8     &  COMP 	& 	$412	^{+69}_{-53}$ 		& $-0.92 \pm 0.07$ 	                                   		                  & 740/479     	& $1.81 \pm 0.07$	\\
	622.7 : 664.7 &COMP 	& 	$371^{+111}_{-71}$ 	& $-1.17 \pm 0.07$ 	                            		                   & 798/477     	& $0.79 \pm 0.04$ 	\\
	838.8 : 1070.2	&  COMP	& 	$278^{+22}_{-18}$ 		& $-1.38 \pm 0.02$                            		        & 1685/360     	& $6.73 \pm 0.09$ 	\\
\hline
\end{tabular}
\end{center}
\smallskip 

\end{table*}

\subsection{Spectral lag analysis}
\begin{figure}
\centering
\includegraphics[width=0.5\textwidth, angle=0]{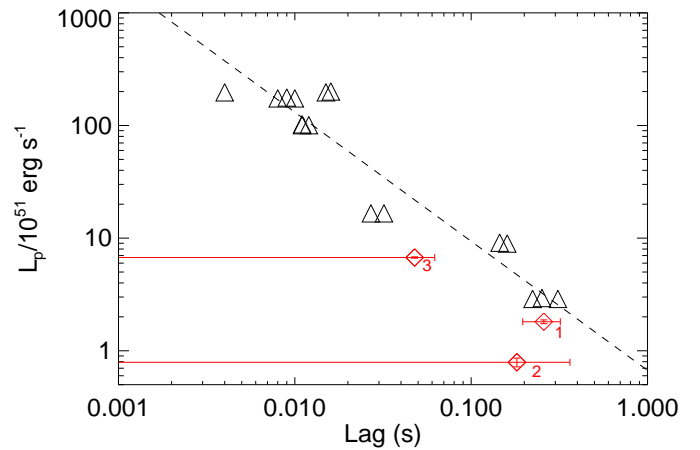}%
\caption{Spectral lag versus peak luminosity (50 keV - 300 keV). Red diamonds indicate the position of the 3 emission episodes of GRB~091024. Black triangles show pulses of other long GRBs presented in \citet{norris00}. }
\label{fig:laglum}
\end{figure}

The spectral lag of a GRB is defined as the time delay between the arrival of high-energy with respect to low-energy gamma-ray photons. Typically, long-duration GRBs exhibit a hard-to-soft spectral evolution due to the decay of the peak spectral energy of the prompt emission over time \citep{ford95, kocevski:2003,hafizi:2007}. This is observed in the energy-resolved GRB light curves as the earlier arrival of emission of a high-energy band relative to a low-energy band. Long GRBs present a large range of spectral lags, with a typical value of 100\,ms~\citep{hakkila:2007}. An anti-correlation between the spectral lags of long GRBs and their peak isotropic luminosities was discovered by \cite{norris:2000}, using a small sample of BATSE/BeppoSAX GRBs and subsequently confirmed with a large sample of \textit{Swift} bursts, albeit with a significant scatter~\citep{ukwatta:2010} (see Fig.~\ref{fig:laglum}). In an extensive study of BATSE GRBs, \citet{hakkila:2008} found that the spectral lag is a property of GRB pulses rather than of the burst itself and can vary significantly for separate pulses of a GRB. The physical origin of the lag is not yet clear. However, it may be a purely kinematic effect, caused by lower-energy high-latitude emission being delayed relative to the line of sight of the observer \citep[e.g.][]{salmonson:2000,ioka:2001}. 

The spectral lag of a GRB is measured by cross-correlating two light curves in different energy bands as a function of temporal lag ~\citep[e.g.][]{norris:2000,foley:2008}. The maximum of the cross-correlation function (CCF) then corresponds to the spectral lag of the GRB. In order to avoid spurious lag measurements due to short-timescale noise variations in the CCF, a function is fit to the CCF and the maximum of the fit to the CCF is taken as the true lag value. Statistical errors on the lag are estimated using a bootstrap method as described in~\cite{norris:2000}. This involves simulating 100 light curves of the GRB by adding Poissonian noise based on the observed count rates to the original light curves and recomputing the lag.

Spectral lags were measured for GRB\,091024 between light curves in the 25--50\,keV and 100--300\,keV energy bands and in the same time intervals as those selected for the spectral analysis. High-time-resolution time-tagged event (TTE) data were used for the first two intervals. TTE data were unavailable for the final emission interval and so CTIME data were used. In each case the lag was computed at a temporal binning of 64\,ms. The results are shown in Table~\ref{table:lags}.

\begin{figure}
\centering
\includegraphics[width=0.5\textwidth, angle=0]{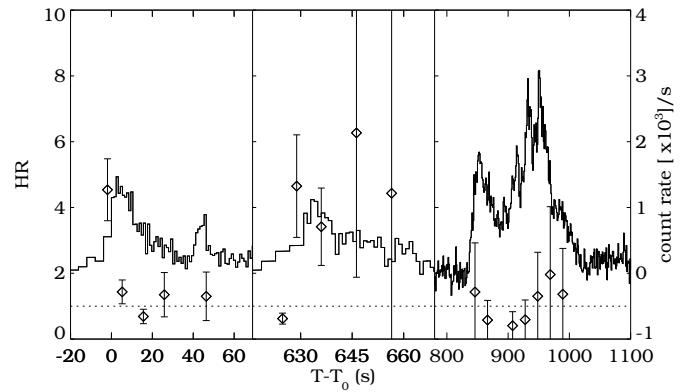}%
\caption{Hardness ratio (HR) of energy bands from (100 keV - 300 keV)/(10 keV - 100 keV). The solid line shows the count light curve of GRB~091024 with a time resolution of 4.096~s immediately before the two GBM triggers and a 1.024~s resolution post-trigger. }
\label{fig:HR}
\end{figure}

 The lag is seen to vary throughout the burst. The positive spectral lag for episode \RM{1} of GRB\,091024 indicates the hard emission leading the soft. 
As argued above, emission episode \RM{1} should therefore show a hard-to-soft evolution. This is in line with the results of the HR analysis (see Fig.~\ref{fig:HR}) and typical for GRB pulses. The lags of episode \RM{2} and \RM{3} are not well constrained. However, both are consistent with zero, i.e. no spectral evolution which is in agreement with Fig.~\ref{fig:HR}.

In Fig.~\ref{fig:laglum} we present the position of the 3 episodes in the lag-luminosity diagram, first presented by \citet{norris00}. According to \citeauthor{norris00} we determined the peak luminosity in the 50 keV to 300 keV energy range and find the position to be consistent with the locations of previous GRBs.

\begin{table}
\renewcommand{\arraystretch}{1.4}
\caption{Results of the spectral lag analysis in the in the 25--50\,keV and 100--300\,keV energy bands. A positive spectral lag indicates the earlier arrival of high-energy photons with respect to low-energy photons.}
\begin{center}
\label{table:lags}
\begin{tabular}{l c l}
\hline
   & $t-t_0$[s] & Spectral Lag [s] \\
\hline
\hline
Episode \RM{1}  & -3.8 - 67.8 & $0.54\pm0.13$ \\
Episode \RM{2}  & 622.7 - 664.7 & $0.38^{+0.38}_{-0.58}$ \\
Episode \RM{3} & 838.8 - 1070.2 & $0.10^{+0.03}_{-0.13}$ \\
\hline
\end{tabular}
\end {center}
\end{table}

\section{Afterglow and optical flash}
\label{sec:afterglow}

\begin{figure}
\centering
\includegraphics[width=0.35\textwidth, angle=270]{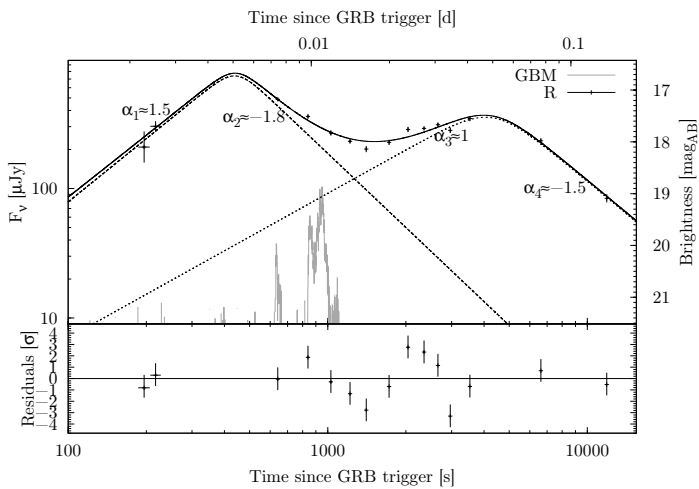}%
\caption{Combined r-band light curve using data points from \citet{henden09, cano09, updike09}. The dashed line shows a reverse shock/forward shock modeling.  The GBM light curve is presented in counts/s to guide the eye. \textit{Swift}-XRT started observing 3000~s after the GBM trigger.}
\label{fig:afterglow}
\end{figure}

Using the data from \citet{henden09, cano09, updike09} the gross behavior of the optical afterglow light curve is shown in Fig.~\ref{fig:afterglow}. 

The light curve has been fitted with the superposition of two different components. Each of these components is represented by a smoothly broken power law \citep{1999A&A...352L..26B}. Given the sparsely sampled data at $t<600$~s, and the strong variability in the light curve there is considerable degeneracy between all fit parameters. We do reach, however, the following firm conclusions: The early optical light curve initially rises and peaks at around 450~s. Forcing the first peak to be simultaneous to the second emission episode in the GBM light curve $F_{\nu}\propto t^{\alpha}$ at $\sim 630$~s results in a worse fit, and would require some fine-tuning of the parameters.  Given that previous observations have shown the optical prompt emission to be quasi-simultaneous or somewhat later than the gamma-ray photons \citep{vestrand05, page06, 2009ApJ...697..758K}, the first optical peak is therefore very likely unrelated to the emission seen in the GBM at 630~s. The initial afterglow rise-index $\alpha_1$ is not well constrained by the data with a value of $\alpha_1 \approx 1.0-2.0$, which is compatible with what has been measured for previous rising afterglow light curves \citep[e.g.][]{molinari07, 2008MNRAS.387..497P, 2009A&A...508..593K, 2009MNRAS.395..490O}. After the first peak, the first light curve component declines with an index of $\alpha_2 \approx -1.8$. This decay would be remarkably steep for a typical pre-jet break afterglow forward shock, but is consistent with the expectation for the decline of a reverse shock \citep{2000ApJ...545..807K}. We point out the possibility that such a steep decline could also be caused by a standard afterglow in a wind-like medium if assuming $p \approx 2.7$. Under the assumption that $p$ remains constant throughout the afterglow, we use the \textit{Swift}-XRT X-ray spectrum\citep{ev07, ev09} between T0+3300 s and T0+50000 s, finding a spectral index of $\beta_{\rm{X}}=0.6\pm0.2$. Using the standard equations \citep[see e.g.][]{zhang06} we infer a value of $p=2.2\pm0.4$ (for $\nu_m<\nu_{\rm{X}}<\nu_c$) which is consistent with a wind-like medium within 90\%. The second peak in the light curve could then be caused for example by a refreshed shock, i.e. a late energy injection into the forward shockwave \citep[e.g.][]{1998ApJ...496L...1R}, or by patchy shells which represent inhomogeneities in the angular energy distribution of the jet \citep[e.g.][]{2000ApJ...535..152K}. However, the interpretation for a wind-like medium would require a much shallower rise \citep[$\alpha \approx 0.5$, see][]{2008MNRAS.387..497P} than the one actually observed ($\alpha=1.5$, see Fig. ~\ref{fig:afterglow}). Although we cannot rule out the possibility of a forward shock in a wind-like circum-burst medium, the light curve evolution argues against this scenario.

After the steep decay, the light curve reaches a temporary minimum at around 1200~s after which it rises again to a second peak at around 4000-5000~s. The light curve coverage is sparse around and after the second peak, but in any case the light curve peaks at a moment when there was no detection of further gamma-ray emission (\textit{Fermi} was in the SAA at this time. However, \textit{Swift}-BAT observed the field of GRB~091024 approximately $3000$~s after the trigger and reported no detection). The second afterglow peak therefore is also unrelated to the prompt gamma-rays. Although sparsely sampled, the rise and the decay indices are consistent with the decay index of a typical afterglow forward shock.

Due to the lack of correlation between the optical and prompt gamma-ray emission, two different processes must have produced the two emission episodes. This is well expected in the internal-external shock model of GRB emission where an external reverse shock arises due to the interaction of the ejecta and the surrounding material. The reverse shock then crosses the emitted shell, thereby accelerating the electrons which then cool adiabatically \citep{sapi99a}. This shock occurs only once, hence emitting a single burst.

There are only a few GRBs where the onset of the optical afterglow could be observed so quickly after the trigger and during the prompt emission. Among these GRBs are 
GRB~990123 \citep{akerlof99, briggs99, sapi99c, soderberg02}, 
GRB~050904 \citep{boer06},
GRB~060124 \citep{romano06}, 
GRB~060418 and GRB 060607A \citep{molinari07},
GRB~061121 \citep{page06}
 and also in GRB~041219A \citep{vestrand05, blake05, mcbreen06}. However, for GRB~041219A, GRB~061121 and GRB~041219A, the optical and prompt gamma emission are highly correlated which indicates a common origin.

We extrapolated the spectrum of the prompt gamma emission of the three emission epochs to calculate the monochromatic flux at a wavelength in the $R_c$-band ($\approx 550\, \rm{nm}$), using

\begin{equation}
F(\nu_0)=\frac{F_{obs}^{[\nu_1,\nu_2]}}{\nu_0^\beta}\,\frac{1-\beta}{\nu_2^{1-\beta}-\nu_1^{1-\beta}}
\end{equation}

where $F_{obs}$ is the energy flux in the frequency range between $\nu_1$ and $\nu_2$, $\nu_0$ is the frequency at 626~nm and $\beta$ is the power law index of 

\begin{equation}
F(\nu_0)=k\nu_0^{\beta},
\end{equation}

i.e. the low-energy photon index of the spectrum, $\alpha$ , minus 1.
For the emission episodes \RM{1} and \RM{2} we used 10~keV and 300~keV for $\nu_1$ and $\nu_2$, respectively. 10~keV to 200~keV was used for episode \RM{3} because the peak energy of the time integrated spectrum is well below 300~keV.
Our estimate of the extrapolated monochromatic flux is conservative in the sense that we took also into consideration the error in $E_{peak}$, i.e. we are using the actual value of $E_{peak}$ minus the 1 sigma error and the shallowest photon index, i.e. $\alpha$ minus 1 sigma for the lower estimate of the monochromatic flux and $E_{peak}+ 1 \sigma$ and the steepest photon index for the upper limit.
The so obtained values for the monochromatic flux density are $40<F(\nu_0) [\mu Jy]<175$, $470<F(\nu_0) [\mu Jy]<2100$ and $14950<F(\nu_0) [\mu Jy]<22450$ for the emission episodes \RM{1}, \RM{2} and \RM{3}, respectively. Before these estimates can be compared directly to the measured flux densities of the afterglow, the latter need to be corrected for Galactic foreground extinction. With $E(B-V)=0.98$~mag and $R(V)=3.1$, yielding $A(V)=3.04$, the $R_c$ band has a corrected magnitude of $A(V)\times0.807=2.34$~mag. Consequently, the intrinsic monochromatic flux of the afterglow is $\sim10$ times \textit{brighter} than the observed one shown in Fig.~\ref{fig:afterglow}. This, in turn, means that the flux densities derived from the extrapolation of the gamma-ray spectrum are considerably lower than the intrinsic afterglow flux densities. Therefore, we can exclude a common prompt gamma-ray and afterglow origin at the times of emission episodes \RM{1} and \RM{2}.

\section{Constraints on the initial Lorentz factor}

We estimate the Lorentz factor at the deceleration time scale using the afterglow peak time following Eq. 1 in \citet{molinari07}. This implicitly assumes that the optical afterglow peak is caused by the fireball forward shock model \citep{sapi99a, mesz06}. The Lorentz factor, for a homogeneous surrounding medium with particle density $n = n_0\, \rm{cm}^{-3}$, at the time of deceleration is then

\begin{equation}
\Gamma_{\rm{dec}} (t_{\rm{peak}}) = \left(\frac{3E_\gamma(1+z)^3}{32 \pi n m_p c^5 \eta t^3_{\rm{peak}}}\right)^{1/8} \approx 160 \left(\frac{E_{\gamma, 53}(1+z)^{3}}{\eta_{0.2} n_0 t^3_{\rm{peak}, 2}}\right)^{1/8}, 
\end{equation}
 
where $E_\gamma = 10^{53}E_{\gamma, 53}$ erg is the isotropic-equivalent energy released by the GRB in the 1~keV to 10~MeV energy band, $\eta = 0.2 \eta_{0.2}$ the radiative efficiency, $t_{\rm{peak}, 2} = t_{\rm{peak}}/(100 \rm{s})$, $m_p$ the proton mass, $c$ the speed of light and $z$ the redshift.
\citet{molinari07} also provide an estimate for a wind environment with

\begin{equation}
\Gamma_{\rm{dec}} (t_{\rm{peak}}) = \left(\frac{E_\gamma (1+z)}{8\pi A m_p c^3 \eta t_{\rm{peak}}}\right)^{1/4},
\end{equation}

where $A$ is a constant defined as $n(r)=Ar^{-2}$ and in this case $A^* = A/(3\times10^{35} \rm{cm}^{-1}) = 1$. The initial Lorentz factor is then estimated by simply multiplying $\Gamma_{dec}$ by 2 \citep{mesz06, pankum00}.
The afterglow light curve peaks twice (Fig.~\ref{fig:afterglow}): the first time at $\sim 400\,\rm{s}$ and a second time at $\sim 4000\,\rm{s}$, respectively. Assuming that the first peak is caused by the forward shock ($t_{\rm{peak}}=400$~s), the Lorentz factor in a homogeneous medium is $\Gamma_0 \approx 290$ and $\Gamma_0 \approx 150$ in a wind environment. However, if one interprets this peak as the optical flash of the reverse shock, as we prefer to do (see Sect. \ref{sec:afterglow}) and uses the peak time of the second optical peak ($t_{\rm{peak}}=4000$~s), one gets $\Gamma_0 \approx 120 $ for the homogeneous medium and $\Gamma_0 \approx 80$ for the wind environment.

\label{sec:lorentz}
According to \citet{lisa01} one can estimate a lower limit of the initial Lorentz factor, $\Gamma_0$, from the spectral properties and the variability timescales of the prompt gamma-ray emission from the GRB.
We used

\begin{equation}
\Gamma_0 > (180/11)^{1/(2\beta+6)}\,  \tau^{1/(\beta+3)}\,  (1+z)^{(\beta -1)/(\beta+3)} \, \rm{, where}
\end{equation}

\begin{equation}
\tau = \frac{\sigma_T\, d_L^2\, (m_ec^2)^{1-\beta}\,f}{c^2\, \delta t\,(\beta-1)}.
\end{equation}

\begin{figure}
\centering
\includegraphics[width=0.5\textwidth, clip]{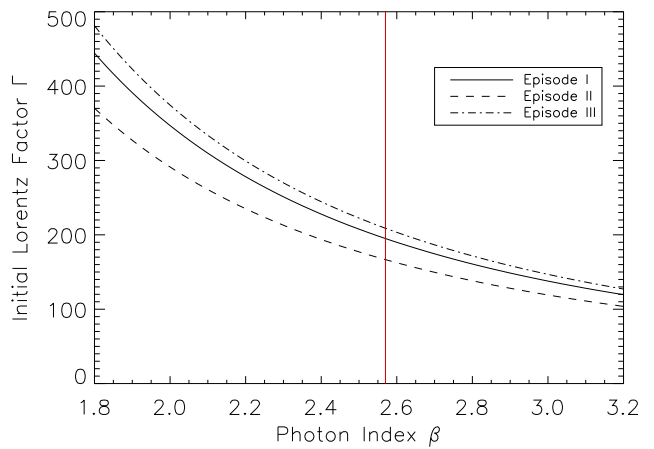}%
\caption{Deduced lower limit of $\Gamma_0$ as a function of the photon index $\beta$ using \citet{lisa01} for episode \RM{1} at $t_0$ (continuous line), episode \RM{2} at $t_0$+630~s (dashed line) and episode \RM{3} at $t_0$+830~s (dashed-dotted line). The red vertical line indicates the photon index of the Band model of the first emission epoch.}
\label{fig:minGamvar}
\end{figure}

In the above equations 
$\delta t$ is the smallest detectable variability and $f$ is the normalization constant of the observed photon flux, defined as $N(E)=f\,E^{-\beta}$. 
Unfortunately, few photons with energies greater than 500 keV were detected. Assuming that the spectrum extends to higher energies we used the high-energy photon index $\beta$ and extrapolated the power-law slope starting at 500~keV into the high-energy domain up to 1~MeV. 
The shortest variability time scale which can be detected in the GBM data changes between the three pulses. Episode \RM{1} has $\delta t = 64\, \rm{ms}$, episode \RM{2} $\delta t = 96\, \rm{ms}$ and episode \RM{3} $\delta t = 32\, \rm{ms}$.
However, only episode \RM{1} has a constrained high-energy photon index $\beta$. Therefore, only for this episode an actual value for $\Gamma_{0, \rm{min}}$ be deduced (intercept between red vertical line and dash-dotted line in Fig.~\ref{fig:minGamvar}) and reported in Table~\ref{tab:minGam}. In Fig.~\ref{fig:minGamvar} we plot the lower limits of the initial Lorentz factors for all three emission episodes as a function of $\beta$.

\begin{table}
\renewcommand{\arraystretch}{1.3}
\caption[ ]{Parameters and lower limit of $\Gamma_0$ obtained using Eqs. 4 and 8 from \citet{lisa01}. The photon flux was determined in the energy range from 500~keV to 1~MeV.}
\label{tab:minGam}
\begin{center}
\begin{tabular}{cccccccc}
\hline
	$\beta$             & flux $F_{ph}$   &  $f$  &   $\Gamma_{0, \rm{min}}$\\
   &                          [ph/cm$^2$/s]  &  [ph/cm$^2$/s/MeV$^{1-\beta}$]  &                           \\
\hline
\hline
$2.57^{-0.39}_{+1.20}$ & 	 $0.056 \pm 0.007$ & $0.044 \pm  0.006$ & $ 195^{+90}_{-110}$\\



\hline
\end{tabular}
\end{center}
\end{table}

\section{Yonetoku and Amati relation}
\label{sec:relations}
\begin{figure}
\centering
\includegraphics[width=0.5\textwidth, angle=0]{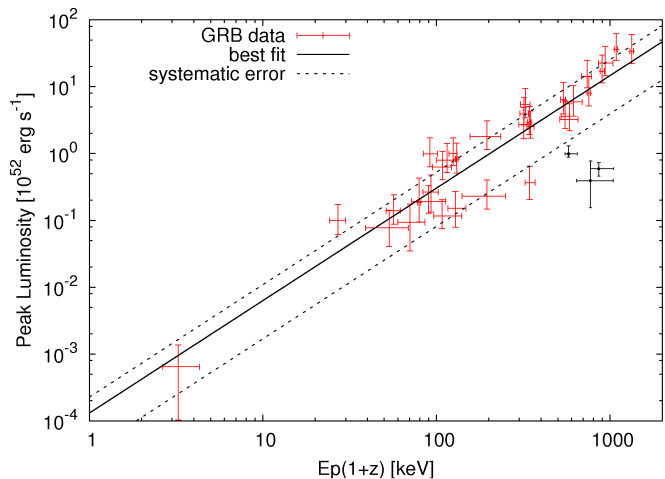}%
\caption{Peak luminosity, $L_p$, and rest frame peak energy, $E_{peak,\rm{rest}}$. Figure adapted from \citet{kodama08}. Red dots show 33 GRBs with $z<1.62$. The solid line shows the fit to the red data points, the dashed lines show the systematic error. Black dots are the values for the three emission epochs in GRB~091024. }
\label{fig:yonetoku}
\end{figure}

\citet{yonetoku04} found a strong correlation between the rest frame peak energy in the $\nu \rm{F}_{\nu}$ spectrum $E_{peak, rest}$  and the 1-s peak luminosity $(L_p)$ in GRBs (so called Yonetoku relation). The peak luminosity is calculated with $L_p=4\pi d_l^2 F_p$, where $d_l$ is the luminosity distance and $F_p$ the peak energy flux. We determined $L_p$, in the energy range between 1 keV and 10 MeV in the rest frame of the GRB for each pulse (though we had to extrapolate above $\sim 500$~keV up to 10 MeV). Interestingly, our values are all below the Yonetoku relation, well outside the one sigma scatter, as is clearly shown by the black dots in Fig.~\ref{fig:yonetoku}. 

\begin{figure}
\centering
\includegraphics[width=0.5\textwidth, angle=0]{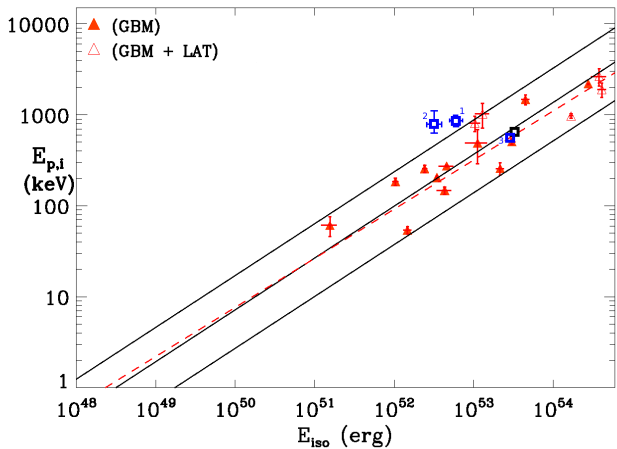}%
\caption{The three emission episodes indicated as blue squares in the $E_{peak,\rm{rest}}$-$E_{iso}$ plane. The black square shows the ``summed'' position of GRB~091024, obtained using time-integrated quantities (see text for details). Red triangles indicate long GBM GRBs as presented in \citet{amati10} together with the best fit to this data (dashed red line). The black solid line shows the best-fit power-law, known as the ``Amati relation'' together with the $\pm2\sigma$ c.l. region \citep{amati08}.}
\label{fig:amati}
\end{figure}

\citet{amati02} found a tight relationship between $E_{peak, rest}$ and the isotropic equivalent bolometric energy,  $E_{iso}$, in the 1~keV to 10~MeV energy range (Amati relation). In order to determine the position of GRB~091024 in the Amati relation, we estimated a time \textit{integrated} $E_{peak}$ value by making a time-weighted $E_{peak}$-average of the 3 emission episodes which results in $E_{peak}=315^{+43}_{-32}$. Summing up $E_{iso}$ gives $3.5\times10^{53}$~erg. 
We present the position of GRB~091024 in the $E_{peak\rm{, rest}}$-$E_{iso}$ plane in Fig.~\ref{fig:amati}.
Additionally, we show the positions of the three emission episodes separately. Similar to GRB~980425 \citep{galama98} and GRB~031203 \citep{malesani04}, emission episodes \RM{1} and \RM{2} are both outliers to the relation. This shows that, in principle, outliers of the Amati relation
could be caused by the sensitivity of the instrument or visibility constraints. If the GRB was active for a longer period than it could actually be observed, it would end up as an outlier on the Amati relation. However, this might then not be due to an intrinsic, physical property of the GRB. Take as an example GRB~091024: \textit{Swift}-BAT only observed the first emission episode because of an Earth-limb constraint. If no other mission would have detected this burst, episode \RM{1} would likely have been interpreted as the complete GRB resulting in a $>2\sigma$ outlier of the Amati relation. 


\begin{table*}
\renewcommand{\arraystretch}{1.3}
\caption[ ]{Rest-frame $E_{peak}$, peak fluxes and peak luminosities of GRB~091024 in various energy ranges. Peak fluxes, $F_p$, are in units of [$10^{-7}$ erg/cm$^2$/s] and peak luminosities, 
$L_p$, in units of [$10^{51}$ erg/s] .}
\label{tab:pfpl}
\begin{center}
\begin{tabular}{cccccccc}
\hline
	            &  $E_{peak}(1+z)$& $F_p$   &   $L_p$                     & $F_p$ & $L_p$                          \\
                       & [keV] & [1 keV - 10 MeV]           &  [1 keV - 10 MeV]  & [50 keV - 300 keV]       &    [50 keV - 300 keV] \\    
\hline
\hline
Episode \RM{1}    & $860^{+140}_{-110}$& $7.0 \pm 1.0$   & $5.3 \pm 0.7$      & $2.1 \pm 0.1$ & $1.6 \pm 0.1$  \\
Episode \RM{2} & $780^{+290}_{-150}$&  $6.2 \pm 3.5$   & $4.6 \pm 2.6$     & $1.0 \pm 0.1$ & $0.7 \pm 0.1$  \\ 
Episode \RM{3}          &$580^{+50}_{-40}$& $14.0 \pm 3.0$ & $10.5 \pm 2.2$   & $2.9 \pm 0.1$ & $2.2 \pm 0.1$ \\
\hline

\end{tabular}
\end{center}
\end{table*}

\section{Summary}
\label{sec:sum}
GRB~0910124 was a very long burst which lasted for $T_{\rm{90}}\approx1020\,\rm{s}$. Soon after the detection by \textit{Fermi}/GBM optical observation started with the Sonoita Research Observatory, the Faulkes North Telescope and the Super-LOTIS telescope. The prompt-ray gamma light curve is characterized by three emission episodes separated by approximately 630~s and 200~s of quiescence, respectively. The optical light curve shows two peaks. The first one occurs well before the onset of the second emission process in gamma-rays. We interpret the first peak as the the reverse shock, thus adding another burst to the sample of GRBs with an optical flash. The second peak, at $t_0$+4000~s, is then due to the forward shock. Afterwards the optical flux declines again.

Using \textit{Fermi}/GBM data, we performed a spectral analysis of the three distinct emission episodes in gamma-rays. Additionally, from the smallest detectable variability time scale, we estimated the lower limit on the bulk Lorentz factor using the Eqs. from \citet{lisa01} and found it to be $\Gamma_{0, \rm{min}} = 195^{+90}_{-110}$. From the peak time of the forward shock, i.e. the time when the blast wave has decelerated, we determined the Lorentz factor to be $\Gamma_0 \approx 120$ according to \citet{molinari07} for a homogeneous ISM which is perfectly consistent with the lower limit mentioned above and with Lorentz factors of other long bursts.

GRB~091024, while being consistent with the Amati relation, is a $> 1 \sigma$ outlier to the Yonetoku relation.

\acknowledgements{The GBM project is supported by the German Bundesministerium f\"ur Wirtschaft und Technologie (BMWi) via the Deutsches Zentrum f\"ur Luft- und Raumfahrt (DLR) under the contract numbers 50 QV 0301 and 50 OG 0502. SF acknowledges the support of the Irish Research Council for Science, Engineering and Technology, cofunded by Marie Curie Actions under FP7.}

\bibliographystyle{aa} 
\bibliography{references} 

\begin{thebibliography}{66}
\expandafter\ifx\csname natexlab\endcsname\relax\def\natexlab#1{#1}\fi

\bibitem[{{Akerlof} {et~al.}(1999){Akerlof}, {Balsano}, {Barthelmy}, {Bloch},
  {Butterworth}, {Casperson}, {Cline}, {Fletcher}, {Frontera}, {Gisler},
  {Heise}, {Hills}, {Kehoe}, {Lee}, {Marshall}, {McKay}, {Miller}, {Piro},
  {Priedhorsky}, {Szymanski}, \& {Wren}}]{akerlof99}
{Akerlof}, C., {Balsano}, R., {Barthelmy}, S., {et~al.} 1999, \nat, 398, 400

\bibitem[{{Amati}(2010)}]{amati10}
{Amati}, L. 2010, ArXiv e-prints

\bibitem[{{Amati} {et~al.}(2002){Amati}, {Frontera}, {Tavani}, {in't Zand},
  {Antonelli}, {Costa}, {Feroci}, {Guidorzi}, {Heise}, {Masetti}, {Montanari},
  {Nicastro}, {Palazzi}, {Pian}, {Piro}, \& {Soffitta}}]{amati02}
{Amati}, L., {Frontera}, F., {Tavani}, M., {et~al.} 2002, \aap, 390, 81

\bibitem[{{Amati} {et~al.}(2008){Amati}, {Guidorzi}, {Frontera}, {Della Valle},
  {Finelli}, {Landi}, \& {Montanari}}]{amati08}
{Amati}, L., {Guidorzi}, C., {Frontera}, F., {et~al.} 2008, \mnras, 391, 577

\bibitem[{{Atwood} {et~al.}(2009){Atwood}, {Abdo}, {Ackermann}, {Althouse},
  {Anderson}, {Axelsson}, {Baldini}, {Ballet}, {Band}, {Barbiellini},
  {Bartelt}, {Bastieri}, {Baughman}, {Bechtol}, {B{\'e}d{\'e}r{\`e}de},
  {Bellardi}, {Bellazzini}, {Berenji}, {Bignami}, {Bisello}, {Bissaldi},
  {Blandford}, {Bloom}, {Bogart}, {Bonamente}, {Bonnell}, {Borgland},
  {Bouvier}, {Bregeon}, {Brez}, {Brigida}, {Bruel}, {Burnett}, {Busetto},
  {Caliandro}, {Cameron}, {Caraveo}, {Carius}, {Carlson}, {Casandjian},
  {Cavazzuti}, {Ceccanti}, {Cecchi}, {Charles}, {Chekhtman}, {Cheung},
  {Chiang}, {Chipaux}, {Cillis}, {Ciprini}, {Claus}, {Cohen-Tanugi},
  {Condamoor}, {Conrad}, {Corbet}, {Corucci}, {Costamante}, {Cutini}, {Davis},
  {Decotigny}, {DeKlotz}, {Dermer}, {de Angelis}, {Digel}, {do Couto e Silva},
  {Drell}, {Dubois}, {Dumora}, {Edmonds}, {Fabiani}, {Farnier}, {Favuzzi},
  {Flath}, {Fleury}, {Focke}, {Funk}, {Fusco}, {Gargano}, {Gasparrini},
  {Gehrels}, {Gentit}, {Germani}, {Giebels}, {Giglietto}, {Giommi}, {Giordano},
  {Glanzman}, {Godfrey}, {Grenier}, {Grondin}, {Grove}, {Guillemot}, {Guiriec},
  {Haller}, {Harding}, {Hart}, {Hays}, {Healey}, {Hirayama}, {Hjalmarsdotter},
  {Horn}, {Hughes}, {J{\'o}hannesson}, {Johansson}, {Johnson}, {Johnson},
  {Johnson}, {Johnson}, {Kamae}, {Katagiri}, {Kataoka}, {Kavelaars}, {Kawai},
  {Kelly}, {Kerr}, {Klamra}, {Kn{\"o}dlseder}, {Kocian}, {Komin}, {Kuehn},
  {Kuss}, {Landriu}, {Latronico}, {Lee}, {Lee}, {Lemoine-Goumard}, {Lionetto},
  {Longo}, {Loparco}, {Lott}, {Lovellette}, {Lubrano}, {Madejski}, {Makeev},
  {Marangelli}, {Massai}, {Mazziotta}, {McEnery}, {Menon}, {Meurer},
  {Michelson}, {Minuti}, {Mirizzi}, {Mitthumsiri}, {Mizuno}, {Moiseev},
  {Monte}, {Monzani}, {Moretti}, {Morselli}, {Moskalenko}, {Murgia},
  {Nakamori}, {Nishino}, {Nolan}, {Norris}, {Nuss}, {Ohno}, {Ohsugi}, {Omodei},
  {Orlando}, {Ormes}, {Paccagnella}, {Paneque}, {Panetta}, {Parent}, {Pearce},
  {Pepe}, {Perazzo}, {Pesce-Rollins}, {Picozza}, {Pieri}, {Pinchera}, {Piron},
  {Porter}, {Poupard}, {Rain{\`o}}, {Rando}, {Rapposelli}, {Razzano}, {Reimer},
  {Reimer}, {Reposeur}, {Reyes}, {Ritz}, {Rochester}, {Rodriguez}, {Romani},
  {Roth}, {Russell}, {Ryde}, {Sabatini}, {Sadrozinski}, {Sanchez}, {Sander},
  {Sapozhnikov}, {Parkinson}, {Scargle}, {Schalk}, {Scolieri}, {Sgr{\`o}},
  {Share}, {Shaw}, {Shimokawabe}, {Shrader}, {Sierpowska-Bartosik}, {Siskind},
  {Smith}, {Smith}, {Spandre}, {Spinelli}, {Starck}, {Stephens}, {Strickman},
  {Strong}, {Suson}, {Tajima}, {Takahashi}, {Takahashi}, {Tanaka}, {Tenze},
  {Tether}, {Thayer}, {Thayer}, {Thompson}, {Tibaldo}, {Tibolla}, {Torres},
  {Tosti}, {Tramacere}, {Turri}, {Usher}, {Vilchez}, {Vitale}, {Wang},
  {Watters}, {Winer}, {Wood}, {Ylinen}, \& {Ziegler}}]{atwood09}
{Atwood}, W.~B., {Abdo}, A.~A., {Ackermann}, M., {et~al.} 2009, \apj, 697, 1071

\bibitem[{{Band} {et~al.}(1993){Band}, {Matteson}, {Ford}, {Schaefer},
  {Palmer}, {Teegarden}, {Cline}, {Briggs}, {Paciesas}, {Pendleton}, {Fishman},
  {Kouveliotou}, {Meegan}, {Wilson}, \& {Lestrade}}]{band93}
{Band}, D., {Matteson}, J., {Ford}, L., {et~al.} 1993, \apj, 413, 281

\bibitem[{{Bennett} {et~al.}(2003){Bennett}, {Hill}, {Hinshaw}, {Nolta},
  {Odegard}, {Page}, {Spergel}, {Weiland}, {Wright}, {Halpern}, {Jarosik},
  {Kogut}, {Limon}, {Meyer}, {Tucker}, \& {Wollack}}]{bennet03}
{Bennett}, C.~L., {Hill}, R.~S., {Hinshaw}, G., {et~al.} 2003, \apjs, 148, 97

\bibitem[{{Beuermann} {et~al.}(1999){Beuermann}, {Hessman}, {Reinsch},
  {Nicklas}, {Vreeswijk}, {Galama}, {Rol}, {van Paradijs}, {Kouveliotou},
  {Frontera}, {Masetti}, {Palazzi}, \& {Pian}}]{1999A&A...352L..26B}
{Beuermann}, K., {Hessman}, F.~V., {Reinsch}, K., {et~al.} 1999, \aap, 352, L26

\bibitem[{{Bissaldi} \& {Connaughton}(2009)}]{bis09}
{Bissaldi}, E. \& {Connaughton}, V. 2009, GRB Coordinates Network, Circular
  Service, 70, 1

\bibitem[{{Blake} {et~al.}(2005){Blake}, {Bloom}, {Starr}, {Falco},
  {Skrutskie}, {Fenimore}, {Duch{\^e}ne}, {Szentgyorgyi}, {Hornstein},
  {Prochaska}, {McCabe}, {Ghez}, {Konopacky}, {Stapelfeldt}, {Hurley},
  {Campbell}, {Kassis}, {Chaffee}, {Gehrels}, {Barthelmy}, {Cummings},
  {Hullinger}, {Krimm}, {Markwardt}, {Palmer}, {Parsons}, {McLean}, \&
  {Tueller}}]{blake05}
{Blake}, C.~H., {Bloom}, J.~S., {Starr}, D.~L., {et~al.} 2005, \nat, 435, 181

\bibitem[{{Bo{\"e}r} {et~al.}(2006){Bo{\"e}r}, {Atteia}, {Damerdji}, {Gendre},
  {Klotz}, \& {Stratta}}]{boer06}
{Bo{\"e}r}, M., {Atteia}, J.~L., {Damerdji}, Y., {et~al.} 2006, \apjl, 638, L71

\bibitem[{{Bouvier} {et~al.}(2009){Bouvier}, {McEnery}, {Chiang}, {Kocevski},
  {Moretti}, {Vasileiou}, \& {Piron}}]{bouvier09}
{Bouvier}, A., {McEnery}, J., {Chiang}, J., {et~al.} 2009, GRB Coordinates
  Network, Circular Service, 114, 1 (2009), 114, 1

\bibitem[{{Briggs} {et~al.}(1999){Briggs}, {Band}, {Kippen}, {Preece},
  {Kouveliotou}, {van Paradijs}, {Share}, {Murphy}, {Matz}, {Connors},
  {Winkler}, {McConnell}, {Ryan}, {Williams}, {Young}, {Dingus}, {Catelli}, \&
  {Wijers}}]{briggs99}
{Briggs}, M.~S., {Band}, D.~L., {Kippen}, R.~M., {et~al.} 1999, \apj, 524, 82

\bibitem[{{Burlon} {et~al.}(2008){Burlon}, {Ghirlanda}, {Ghisellini},
  {Lazzati}, {Nava}, {Nardini}, \& {Celotti}}]{burlon08}
{Burlon}, D., {Ghirlanda}, G., {Ghisellini}, G., {et~al.} 2008, \apjl, 685, L19

\bibitem[{{Campana} {et~al.}(2006){Campana}, {Mangano}, {Blustin}, {Brown},
  {Burrows}, {Chincarini}, {Cummings}, {Cusumano}, {Della Valle}, {Malesani},
  {M{\'e}sz{\'a}ros}, {Nousek}, {Page}, {Sakamoto}, {Waxman}, {Zhang}, {Dai},
  {Gehrels}, {Immler}, {Marshall}, {Mason}, {Moretti}, {O'Brien}, {Osborne},
  {Page}, {Romano}, {Roming}, {Tagliaferri}, {Cominsky}, {Giommi}, {Godet},
  {Kennea}, {Krimm}, {Angelini}, {Barthelmy}, {Boyd}, {Palmer}, {Wells}, \&
  {White}}]{campana06}
{Campana}, S., {Mangano}, V., {Blustin}, A.~J., {et~al.} 2006, \nat, 442, 1008

\bibitem[{{Cano} {et~al.}(2009){Cano}, {Guidorzi}, {Mundell}, {Bersier},
  {Clay}, {Kobayashi}, {Melandri}, {Mottram}, {Smith}, {Steele}, {Gomboc},
  {O'Brien}, {Tanvir}, \& {Marshall}}]{cano09}
{Cano}, Z., {Guidorzi}, C., {Mundell}, C.~G., {et~al.} 2009, GRB Coordinates
  Network, Circular Service, 66, 1

\bibitem[{{Cenko} {et~al.}(2009){Cenko}, {Kasliwal}, \& {Kulkarni}}]{cenko09}
{Cenko}, S.~B., {Kasliwal}, M.~M., \& {Kulkarni}, S.~R. 2009, GRB Coordinates
  Network, Circular Service, 93, 1

\bibitem[{{Cucchiara} {et~al.}(2009){Cucchiara}, {Fox}, \& {Tanvir}}]{cucc09}
{Cucchiara}, A., {Fox}, D., \& {Tanvir}, N. 2009, GRB Coordinates Network,
  Circular Service, 65, 1

\bibitem[{{Evans} {et~al.}(2009){Evans}, {Beardmore}, {Page}, {Osborne},
  {O'Brien}, {Willingale}, {Starling}, {Burrows}, {Godet}, {Vetere}, {Racusin},
  {Goad}, {Wiersema}, {Angelini}, {Capalbi}, {Chincarini}, {Gehrels}, {Kennea},
  {Margutti}, {Morris}, {Mountford}, {Pagani}, {Perri}, {Romano}, \&
  {Tanvir}}]{ev09}
{Evans}, P.~A., {Beardmore}, A.~P., {Page}, K.~L., {et~al.} 2009, \mnras, 397,
  1177

\bibitem[{{Evans} {et~al.}(2007){Evans}, {Beardmore}, {Page}, {Tyler},
  {Osborne}, {Goad}, {O'Brien}, {Vetere}, {Racusin}, {Morris}, {Burrows},
  {Capalbi}, {Perri}, {Gehrels}, \& {Romano}}]{ev07}
{Evans}, P.~A., {Beardmore}, A.~P., {Page}, K.~L., {et~al.} 2007, \aap, 469,
  379

\bibitem[{{Foley} {et~al.}(2008){Foley}, {McGlynn}, {Hanlon}, {McBreen}, \&
  {McBreen}}]{foley:2008}
{Foley}, S., {McGlynn}, S., {Hanlon}, L., {McBreen}, S., \& {McBreen}, B. 2008,
  \aap, 484, 143

\bibitem[{{Ford} {et~al.}(1995){Ford}, {Band}, {Matteson}, {Briggs},
  {Pendleton}, {Preece}, {Paciesas}, {Teegarden}, {Palmer}, {Schaefer},
  {Cline}, {Fishman}, {Kouveliotou}, {Meegan}, {Wilson}, \&
  {Lestrade}}]{ford95}
{Ford}, L.~A., {Band}, D.~L., {Matteson}, J.~L., {et~al.} 1995, \apj, 439, 307

\bibitem[{{Galama} {et~al.}(1998){Galama}, {Vreeswijk}, {van Paradijs},
  {Kouveliotou}, {Augusteijn}, {B{\"o}hnhardt}, {Brewer}, {Doublier},
  {Gonzalez}, {Leibundgut}, {Lidman}, {Hainaut}, {Patat}, {Heise}, {in't Zand},
  {Hurley}, {Groot}, {Strom}, {Mazzali}, {Iwamoto}, {Nomoto}, {Umeda},
  {Nakamura}, {Young}, {Suzuki}, {Shigeyama}, {Koshut}, {Kippen}, {Robinson},
  {de Wildt}, {Wijers}, {Tanvir}, {Greiner}, {Pian}, {Palazzi}, {Frontera},
  {Masetti}, {Nicastro}, {Feroci}, {Costa}, {Piro}, {Peterson}, {Tinney},
  {Boyle}, {Cannon}, {Stathakis}, {Sadler}, {Begam}, \& {Ianna}}]{galama98}
{Galama}, T.~J., {Vreeswijk}, P.~M., {van Paradijs}, J., {et~al.} 1998, \nat,
  395, 670

\bibitem[{{Giblin} {et~al.}(2002){Giblin}, {Connaughton}, {van Paradijs},
  {Preece}, {Briggs}, {Kouveliotou}, {Wijers}, \& {Fishman}}]{giblin02}
{Giblin}, T.~W., {Connaughton}, V., {van Paradijs}, J., {et~al.} 2002, \apj,
  570, 573

\bibitem[{{Golenetskii} {et~al.}(2009){Golenetskii}, {Aptekar}, {Mazets},
  {Pal'Shin}, {Frederiks}, {Oleynik}, {Ulanov}, {Svinkin}, \& {Cline}}]{gol09}
{Golenetskii}, S., {Aptekar}, R., {Mazets}, E., {et~al.} 2009, GRB Coordinates
  Network, Circular Service, 83, 1

\bibitem[{{Hafizi} \& {Mochkovitch}(2007)}]{hafizi:2007}
{Hafizi}, M. \& {Mochkovitch}, R. 2007, \aap, 465, 67

\bibitem[{{Hakkila} {et~al.}(2008){Hakkila}, {Giblin}, {Norris}, {Fragile}, \&
  {Bonnell}}]{hakkila:2008}
{Hakkila}, J., {Giblin}, T.~W., {Norris}, J.~P., {Fragile}, P.~C., \&
  {Bonnell}, J.~T. 2008, \apjl, 677, L81

\bibitem[{{Hakkila} {et~al.}(2007){Hakkila}, {Giblin}, {Young}, {Fuller},
  {Peters}, {Nolan}, {Sonnett}, {Haglin}, \& {Roiger}}]{hakkila:2007}
{Hakkila}, J., {Giblin}, T.~W., {Young}, K.~C., {et~al.} 2007, \apjs, 169, 62

\bibitem[{{Henden} {et~al.}(2009){Henden}, {Gross}, {Denny}, {Terrell}, \&
  {Cooney}}]{henden09}
{Henden}, A., {Gross}, J., {Denny}, B., {Terrell}, D., \& {Cooney}, W. 2009,
  GRB Coordinates Network, Circular Service, 73, 1

\bibitem[{{Ioka} \& {Nakamura}(2001)}]{ioka:2001}
{Ioka}, K. \& {Nakamura}, T. 2001, \apjl, 554, L163

\bibitem[{{Klebesadel} {et~al.}(1973){Klebesadel}, {Strong}, \&
  {Olson}}]{kleb73}
{Klebesadel}, R.~W., {Strong}, I.~B., \& {Olson}, R.~A. 1973, \apjl, 182, L85

\bibitem[{{Kobayashi}(2000)}]{2000ApJ...545..807K}
{Kobayashi}, S. 2000, \apj, 545, 807

\bibitem[{{Kocevski} \& {Liang}(2003)}]{kocevski:2003}
{Kocevski}, D. \& {Liang}, E. 2003, \apj, 594, 385

\bibitem[{{Kodama} {et~al.}(2008){Kodama}, {Yonetoku}, {Murakami}, {Tanabe},
  {Tsutsui}, \& {Nakamura}}]{kodama08}
{Kodama}, Y., {Yonetoku}, D., {Murakami}, T., {et~al.} 2008, \mnras, 391, L1

\bibitem[{{Kr{\"u}hler} {et~al.}(2009{\natexlab{a}}){Kr{\"u}hler}, {Greiner},
  {Afonso}, {Burlon}, {Clemens}, {Filgas}, {Kann}, {Klose}, {K{\"u}pc{\"u}
  Yolda{\c s}}, {McBreen}, {Olivares}, {Rau}, {Rossi}, {Schulze}, {Szokoly},
  {Updike}, \& {Yolda{\c s}}}]{2009A&A...508..593K}
{Kr{\"u}hler}, T., {Greiner}, J., {Afonso}, P., {et~al.} 2009{\natexlab{a}},
  \aap, 508, 593

\bibitem[{{Kr{\"u}hler} {et~al.}(2009{\natexlab{b}}){Kr{\"u}hler}, {Greiner},
  {McBreen}, {Klose}, {Rossi}, {Afonso}, {Clemens}, {Filgas}, {Yolda{\c s}},
  {Szokoly}, \& {Yolda{\c s}}}]{2009ApJ...697..758K}
{Kr{\"u}hler}, T., {Greiner}, J., {McBreen}, S., {et~al.} 2009{\natexlab{b}},
  \apj, 697, 758

\bibitem[{{Kumar} \& {Piran}(2000)}]{2000ApJ...535..152K}
{Kumar}, P. \& {Piran}, T. 2000, \apj, 535, 152

\bibitem[{{Lithwick} \& {Sari}(2001)}]{lisa01}
{Lithwick}, Y. \& {Sari}, R. 2001, \apj, 555, 540

\bibitem[{{Malesani} {et~al.}(2004){Malesani}, {Tagliaferri}, {Chincarini},
  {Covino}, {Della Valle}, {Fugazza}, {Mazzali}, {Zerbi}, {D'Avanzo},
  {Kalogerakos}, {Simoncelli}, {Antonelli}, {Burderi}, {Campana}, {Cucchiara},
  {Fiore}, {Ghirlanda}, {Goldoni}, {G{\"o}tz}, {Mereghetti}, {Mirabel},
  {Romano}, {Stella}, {Minezaki}, {Yoshii}, \& {Nomoto}}]{malesani04}
{Malesani}, D., {Tagliaferri}, G., {Chincarini}, G., {et~al.} 2004, \apjl, 609,
  L5

\bibitem[{{Marshall} {et~al.}(2009){Marshall}, {Baumgartner}, {Beardmore},
  {Burrows}, {Campana}, {Evans}, {Holland}, {Hoversten}, {Markwardt},
  {O'Brien}, {Page}, {Palmer}, {Sakamoto}, {Sbarufatti}, {Starling}, \&
  {Ukwatta}}]{mar09}
{Marshall}, F.~E., {Baumgartner}, W.~H., {Beardmore}, A.~P., {et~al.} 2009, GRB
  Coordinates Network, Circular Service, 62, 1

\bibitem[{{McBreen} {et~al.}(2006){McBreen}, {Hanlon}, {McGlynn}, {McBreen},
  {Foley}, {Preece}, {von Kienlin}, \& {Williams}}]{mcbreen06}
{McBreen}, S., {Hanlon}, L., {McGlynn}, S., {et~al.} 2006, \aap, 455, 433

\bibitem[{{Meegan} {et~al.}(2009){Meegan}, {Lichti}, {Bhat}, {Bissaldi},
  {Briggs}, {Connaughton}, {Diehl}, {Fishman}, {Greiner}, {Hoover}, {van der
  Horst}, {von Kienlin}, {Kippen}, {Kouveliotou}, {McBreen}, {Paciesas},
  {Preece}, {Steinle}, {Wallace}, {Wilson}, \& {Wilson-Hodge}}]{meegan09}
{Meegan}, C., {Lichti}, G., {Bhat}, P.~N., {et~al.} 2009, \apj, 702, 791

\bibitem[{{M{\'e}sz{\'a}ros}(2006)}]{mesz06}
{M{\'e}sz{\'a}ros}, P. 2006, Rep. Prog. Phys., 69, 2259

\bibitem[{{Molinari} {et~al.}(2007){Molinari}, {Vergani}, {Malesani}, {Covino},
  {D'Avanzo}, {Chincarini}, {Zerbi}, {Antonelli}, {Conconi}, {Testa}, {Tosti},
  {Vitali}, {D'Alessio}, {Malaspina}, {Nicastro}, {Palazzi}, {Guetta},
  {Campana}, {Goldoni}, {Masetti}, {Meurs}, {Monfardini}, {Norci}, {Pian},
  {Piranomonte}, {Rizzuto}, {Stefanon}, {Stella}, {Tagliaferri}, {Ward},
  {Ihle}, {Gonzalez}, {Pizarro}, {Sinclaire}, \& {Valenzuela}}]{molinari07}
{Molinari}, E., {Vergani}, S.~D., {Malesani}, D., {et~al.} 2007, \aap, 469, L13

\bibitem[{{Nicastro} {et~al.}(2004){Nicastro}, {in't Zand}, {Amati},
  {Golenetskii}, {Castro-Tirado}, {Gorosabel}, {Lazzati}, {Costa}, {De
  Pasquale}, {Feroci}, {Heise}, {Pian}, {Piro}, {S{\'a}nchez-Fern{\'a}ndez}, \&
  {Tristram}}]{nicastro04}
{Nicastro}, L., {in't Zand}, J.~J.~M., {Amati}, L., {et~al.} 2004, \aap, 427,
  445

\bibitem[{{Norris} {et~al.}(2000{\natexlab{a}}){Norris}, {Marani}, \&
  {Bonnell}}]{norris:2000}
{Norris}, J.~P., {Marani}, G.~F., \& {Bonnell}, J.~T. 2000{\natexlab{a}}, \apj,
  534, 248

\bibitem[{{Norris} {et~al.}(2000{\natexlab{b}}){Norris}, {Marani}, \&
  {Bonnell}}]{norris00}
{Norris}, J.~P., {Marani}, G.~F., \& {Bonnell}, J.~T. 2000{\natexlab{b}}, \apj,
  534, 248

\bibitem[{{Oates} {et~al.}(2009){Oates}, {Page}, {Schady}, {de Pasquale},
  {Koch}, {Breeveld}, {Brown}, {Chester}, {Holland}, {Hoversten}, {Kuin},
  {Marshall}, {Roming}, {Still}, {vanden Berk}, {Zane}, \&
  {Nousek}}]{2009MNRAS.395..490O}
{Oates}, S.~R., {Page}, M.~J., {Schady}, P., {et~al.} 2009, \mnras, 395, 490

\bibitem[{{Page} \& {Marshall}(2009)}]{page09}
{Page}, K.~L. \& {Marshall}, F.~E. 2009, GRB Coordinates Network, Circular
  Service, 69, 1

\bibitem[{{Page} {et~al.}(2007){Page}, {Willingale}, {Osborne}, {Zhang},
  {Godet}, {Marshall}, {Melandri}, {Norris}, {O'Brien}, {Pal'shin}, {Rol},
  {Romano}, {Starling}, {Schady}, {Yost}, {Barthelmy}, {Beardmore}, {Cusumano},
  {Burrows}, {De Pasquale}, {Ehle}, {Evans}, {Gehrels}, {Goad}, {Golenetskii},
  {Guidorzi}, {Mundell}, {Page}, {Ricker}, {Sakamoto}, {Schaefer},
  {Stamatikos}, {Troja}, {Ulanov}, {Yuan}, \& {Ziaeepour}}]{page06}
{Page}, K.~L., {Willingale}, R., {Osborne}, J.~P., {et~al.} 2007, \apj, 663,
  1125

\bibitem[{{Pal'Shin} {et~al.}(2008){Pal'Shin}, {Aptekar}, {Frederiks},
  {Golenetskii}, {Il'Inskii}, {Mazets}, {Yamaoka}, {Ohno}, {Hurley},
  {Sakamoto}, {Oleynik}, {Ulanov}, {Mitrofanov}, {Golovin}, {Lirvak}, {Sanin},
  {Boynton}, {Fellows}, {Harshman}, {Shinohara}, \& {Starr}}]{palshin08}
{Pal'Shin}, V., {Aptekar}, R., {Frederiks}, D., {et~al.} 2008, in American
  Institute of Physics Conference Series, Vol. 1000, American Institute of
  Physics Conference Series, ed. {M.~Galassi, D.~Palmer, \& E.~Fenimore},
  117--120

\bibitem[{{Panaitescu} \& {Kumar}(2000)}]{pankum00}
{Panaitescu}, A. \& {Kumar}, P. 2000, \apj, 543, 66

\bibitem[{{Panaitescu} \& {Vestrand}(2008)}]{2008MNRAS.387..497P}
{Panaitescu}, A. \& {Vestrand}, W.~T. 2008, \mnras, 387, 497

\bibitem[{{Racusin} {et~al.}(2008){Racusin}, {Karpov}, {Sokolowski}, {Granot},
  {Wu}, {Pal'Shin}, {Covino}, {van der Horst}, {Oates}, {Schady}, {Smith},
  {Cummings}, {Starling}, {Piotrowski}, {Zhang}, {Evans}, {Holland}, {Malek},
  {Page}, {Vetere}, {Margutti}, {Guidorzi}, {Kamble}, {Curran}, {Beardmore},
  {Kouveliotou}, {Mankiewicz}, {Melandri}, {O'Brien}, {Page}, {Piran},
  {Tanvir}, {Wrochna}, {Aptekar}, {Barthelmy}, {Bartolini}, {Beskin}, {Bondar},
  {Bremer}, {Campana}, {Castro-Tirado}, {Cucchiara}, {Cwiok}, {D'Avanzo},
  {D'Elia}, {Della Valle}, {de Ugarte Postigo}, {Dominik}, {Falcone}, {Fiore},
  {Fox}, {Frederiks}, {Fruchter}, {Fugazza}, {Garrett}, {Gehrels},
  {Golenetskii}, {Gomboc}, {Gorosabel}, {Greco}, {Guarnieri}, {Immler},
  {Jelinek}, {Kasprowicz}, {La Parola}, {Levan}, {Mangano}, {Mazets},
  {Molinari}, {Moretti}, {Nawrocki}, {Oleynik}, {Osborne}, {Pagani}, {Pandey},
  {Paragi}, {Perri}, {Piccioni}, {Ramirez-Ruiz}, {Roming}, {Steele}, {Strom},
  {Testa}, {Tosti}, {Ulanov}, {Wiersema}, {Wijers}, {Winters}, {Zarnecki},
  {Zerbi}, {M{\'e}sz{\'a}ros}, {Chincarini}, \& {Burrows}}]{racusin08}
{Racusin}, J.~L., {Karpov}, S.~V., {Sokolowski}, M., {et~al.} 2008, \nat, 455,
  183

\bibitem[{{Rees} \& {Meszaros}(1998)}]{1998ApJ...496L...1R}
{Rees}, M.~J. \& {Meszaros}, P. 1998, \apjl, 496, L1+

\bibitem[{{Romano} {et~al.}(2006){Romano}, {Campana}, {Chincarini}, {Cummings},
  {Cusumano}, {Holland}, {Mangano}, {Mineo}, {Page}, {Pal'Shin}, {Rol},
  {Sakamoto}, {Zhang}, {Aptekar}, {Barbier}, {Barthelmy}, {Beardmore}, {Boyd},
  {Burrows}, {Capalbi}, {Fenimore}, {Frederiks}, {Gehrels}, {Giommi}, {Goad},
  {Godet}, {Golenetskii}, {Guetta}, {Kennea}, {La Parola}, {Malesani},
  {Marshall}, {Moretti}, {Nousek}, {O'Brien}, {Osborne}, {Perri}, \&
  {Tagliaferri}}]{romano06}
{Romano}, P., {Campana}, S., {Chincarini}, G., {et~al.} 2006, \aap, 456, 917

\bibitem[{{Salmonson}(2000)}]{salmonson:2000}
{Salmonson}, J.~D. 2000, \apjl, 544, L115

\bibitem[{{Sari} \& {Piran}(1999{\natexlab{a}})}]{sapi99c}
{Sari}, R. \& {Piran}, T. 1999{\natexlab{a}}, \apjl, 517, L109

\bibitem[{{Sari} \& {Piran}(1999{\natexlab{b}})}]{sapi99a}
{Sari}, R. \& {Piran}, T. 1999{\natexlab{b}}, \apj, 520, 641

\bibitem[{{Soderberg} \& {Ramirez-Ruiz}(2002)}]{soderberg02}
{Soderberg}, A.~M. \& {Ramirez-Ruiz}, E. 2002, \mnras, 330, L24

\bibitem[{{Spergel} {et~al.}(2003){Spergel}, {Verde}, {Peiris}, {Komatsu},
  {Nolta}, {Bennett}, {Halpern}, {Hinshaw}, {Jarosik}, {Kogut}, {Limon},
  {Meyer}, {Page}, {Tucker}, {Weiland}, {Wollack}, \& {Wright}}]{spergel03}
{Spergel}, D.~N., {Verde}, L., {Peiris}, H.~V., {et~al.} 2003, \apjs, 148, 175

\bibitem[{{Ukwatta} {et~al.}(2010){Ukwatta}, {Stamatikos}, {Dhuga}, {Sakamoto},
  {Barthelmy}, {Eskandarian}, {Gehrels}, {Maximon}, {Norris}, \&
  {Parke}}]{ukwatta:2010}
{Ukwatta}, T.~N., {Stamatikos}, M., {Dhuga}, K.~S., {et~al.} 2010, \apj, 711,
  1073

\bibitem[{{Updike} {et~al.}(2009){Updike}, {Hartmann}, {Milne}, \&
  {Williams}}]{updike09}
{Updike}, A.~C., {Hartmann}, D.~H., {Milne}, P.~A., \& {Williams}, G.~G. 2009,
  GRB Coordinates Network, Circular Service, 74, 1 (2009), 74, 1

\bibitem[{{Vestrand} {et~al.}(2005){Vestrand}, {Wozniak}, {Wren}, {Fenimore},
  {Sakamoto}, {White}, {Casperson}, {Davis}, {Evans}, {Galassi}, {McGowan},
  {Schier}, {Asa}, {Barthelmy}, {Cummings}, {Gehrels}, {Hullinger}, {Krimm},
  {Markwardt}, {McLean}, {Palmer}, {Parsons}, \& {Tueller}}]{vestrand05}
{Vestrand}, W.~T., {Wozniak}, P.~R., {Wren}, J.~A., {et~al.} 2005, \nat, 435,
  178

\bibitem[{{Yonetoku} {et~al.}(2004){Yonetoku}, {Murakami}, {Nakamura},
  {Yamazaki}, {Inoue}, \& {Ioka}}]{yonetoku04}
{Yonetoku}, D., {Murakami}, T., {Nakamura}, T., {et~al.} 2004, \apj, 609, 935

\bibitem[{{Zhang} {et~al.}(2006){Zhang}, {Fan}, {Dyks}, {Kobayashi},
  {M{\'e}sz{\'a}ros}, {Burrows}, {Nousek}, \& {Gehrels}}]{zhang06}
{Zhang}, B., {Fan}, Y.~Z., {Dyks}, J., {et~al.} 2006, \apj, 642, 354

\end{thebibliography}

\end{document}